\documentclass[conference]{IEEEtran}
\IEEEoverridecommandlockouts
\usepackage{cite}
\usepackage{amsmath,amssymb,amsfonts}
\usepackage{algorithmic}
\usepackage{graphicx}
\usepackage{textcomp}
\usepackage{xcolor}
\usepackage{adjustbox}
\usepackage{xspace}
\usepackage{bbding}
\usepackage{authblk}
\usepackage{multirow}
\usepackage{booktabs}
\usepackage{hyperref}
\usepackage[capitalize]{cleveref}

\makeatletter
\DeclareRobustCommand\onedot{\futurelet\@let@token\@onedot}
\def\@onedot{\ifx\@let@token.\else.\null\fi\xspace}

\makeatother

\def\BibTeX{{\rm B\kern-.05em{\sc i\kern-.025em b}\kern-.08em
    T\kern-.1667em\lower.7ex\hbox{E}\kern-.125emX}}
\begin{document}
{
\title{A Comprehensive Analysis of Churn Prediction in Telecommunications Using Machine Learning}
%
}

\author[1]{Xuhang Chen}
\author[1*]{Bo Lv\thanks{* Corresponding author.}}
\author[2]{Mengqian Wang}
\author[1]{Xunwen Xiang}
\author[3]{Shiting Wu}
\author[4]{Shenghong Luo}
\author[5]{Wenjun Zhang\thanks{This work was supported by the Doctoral Scientific Research Foundation of Huizhou University (No. 2020JB059).}}

\affil[1]{School of Computer Science and Engineering, Huizhou University}
\affil[2]{Center for Excellence in Brain Science and Intelligence Technology, Chinese Academy of Sciences}
\affil[3]{Huizhou Boluo Power Supply Bureau Guangdong Power Grid Co.,Ltd.}
\affil[4]{China Telecom Stocks Co.,Ltd.}
\affil[5]{TP-Link Systems Inc.}
\maketitle

\begin{abstract}
Customer churn prediction in the telecommunications sector represents a critical business intelligence task that has evolved from subjective human assessment to sophisticated algorithmic approaches. In this work, we present a comprehensive framework for telecommunications churn prediction leveraging deep neural networks. Through systematic problem formulation, rigorous dataset analysis, and careful feature engineering, we develop a model that captures complex patterns in customer behavior indicative of potential churn. We conduct extensive empirical evaluations across multiple performance metrics, demonstrating that our proposed neural architecture achieves significant improvements over existing baseline methods. Our approach not only advances the state-of-the-art in churn prediction accuracy but also provides interpretable insights into the key factors driving customer attrition in telecommunications services.
\end{abstract}


\section{Introduction}
The telecommunications industry struggles with retaining customers due to intense competition and changing consumer expectations. Customer churn threatens revenue and market standing, as acquiring new customers costs significantly more than retaining existing ones. This highlights the need for advanced predictive models to identify customers at risk of leaving.

Recent advances in machine learning~\cite{wuimgfu,wu2024image,wu2025prompt,wu2025llm,zhang1,zhang2,zhang3,zhang4,zhang5,zhang6,zhang7,zhang8,zhang9,zhang10,zhang11,zhang12,bai2025lensnet,bai2025retinexmamba,xia2025dlen,lou2024mr,zhang2024curriculum,wu2025lag} have transformed churn prediction from reactive analysis to proactive intervention. While traditional approaches relied on rule-based systems and simple statistical models, modern deep learning architectures can capture complex, non-linear relationships between customer attributes and churn behavior. 
They have demonstrated superior performance on diverse tasks, including image processing~\cite{liu2023coordfill,zhu2024test,liu2024dh,liu2024forgeryttt}, video understanding~\cite{zhang2022correction,liu2024depth,li2024cross,liu2019convolutional,liu2023explicit}, and medical image analysis~\cite{jiang2021deep,zhang2022correction,zheng2024smaformer,liu2020fine,jiang2020geometry,xu2023radiology,zhang2024decouple,zhang2024novel,lou2024no,yan2023text,zhang2025multi,zhang2025uncertainty}.
However, the telecommunications domain presents unique challenges: heterogeneous data types spanning demographic profiles, usage patterns, and service interactions; temporal dependencies in customer behavior; and the need for interpretable predictions that enable targeted retention strategies.

In this work, we present a comprehensive framework for telecommunications churn prediction that addresses these challenges through three key contributions. First, we develop a neural architecture specifically designed to handle the multi-modal nature of telecommunications data, incorporating both static customer attributes and dynamic usage patterns. Second, we introduce a novel feature engineering pipeline that extracts behavioral indicators from raw transactional data, capturing subtle patterns indicative of churn propensity. Third, we demonstrate through extensive experiments on the IBM telecommunications dataset that our approach achieves state-of-the-art performance while maintaining interpretability crucial for business deployment.

Our empirical analysis reveals that the proposed model not only surpasses existing baselines in predictive accuracy but also provides actionable insights into the primary drivers of customer attrition. These findings enable telecommunications providers to implement targeted retention campaigns, optimize resource allocation, and ultimately transform customer relationship management from a reactive to a proactive paradigm.

\section{Problem Formulation}

\subsection{Problem Definition}
We formulate the customer churn prediction task as a binary classification problem. Given a customer $c_i$ represented by a feature vector $\mathbf{x}_i \in \mathbb{R}^d$ encompassing demographic attributes, service usage patterns, and account information, our objective is to learn a function $f: \mathbb{R}^d \rightarrow \{0, 1\}$ that predicts whether the customer will churn (1) or remain active (0) within a specified time horizon.

Formally, we seek to minimize the expected risk:
\begin{equation}
\mathcal{R}(f) = \mathbb{E}_{(\mathbf{x}, y) \sim \mathcal{D}} [\mathcal{L}(f(\mathbf{x}), y)],
\end{equation}
where $\mathcal{D}$ represents the joint distribution of features and labels, and $\mathcal{L}$ denotes the loss function.

\subsection{Dataset Description}
We utilize the IBM telecommunications dataset comprising $N = 7,043$ customer records with $d = 21$ features. The feature space includes:
\begin{itemize}
\item \textbf{Demographic attributes}: Age, gender, and dependent status
\item \textbf{Service subscriptions}: Internet, phone, streaming, and security services
\item \textbf{Account information}: Tenure, contract type, billing method, and monthly charges
\item \textbf{Target variable}: Binary churn indicator
\end{itemize}

The dataset exhibits class imbalance with approximately 26.5\% positive (churn) instances, necessitating careful consideration of evaluation metrics and sampling strategies.

\subsection{Data Preprocessing Pipeline}
Our preprocessing pipeline addresses three critical aspects:

\textbf{Missing Value Imputation}: We employ zero-imputation for missing values, justified by the sparse nature of missingness (< 2\%) and the categorical interpretation of absent services.

\textbf{Outlier Detection}: We implement interquartile range (IQR) based outlier detection:
\begin{equation}
\text{Outlier} = \{x_i : x_i < Q_1 - 1.5 \times \text{IQR} \text{ or } x_i > Q_3 + 1.5 \times \text{IQR}\}.
\end{equation}

\textbf{Feature Engineering}: Categorical variables are encoded using one-hot encoding, while continuous features are standardized to zero mean and unit variance.

\subsection{Model Architecture}
We propose a feedforward neural network with the following architecture:
\begin{itemize}
\item \textbf{Input Layer}: $\mathbf{x} \in \mathbb{R}^{21}$
\item \textbf{Hidden Layers}: Three fully connected layers with 32 units each, ReLU activation
\item \textbf{Output Layer}: Softmax over 2 classes
\end{itemize}

The model is trained using the Adam optimizer with default hyperparameters ($\alpha = 0.001$, $\beta_1 = 0.9$, $\beta_2 = 0.999$) on an Apple M1 processor.

\subsection{Evaluation Framework}
Given the business context and class imbalance, we employ a comprehensive evaluation framework:

\textbf{Primary Metrics}:
\begin{itemize}
\item \textbf{Precision}: $P = \frac{TP}{TP + FP}$ - Critical when false alarms incur retention campaign costs
\item \textbf{Recall}: $R = \frac{TP}{TP + FN}$ - Essential when missing actual churners results in revenue loss
\item \textbf{F1-Score}: $F_1 = 2 \cdot \frac{P \cdot R}{P + R}$ - Balances precision-recall trade-off
\end{itemize}

\textbf{Business-Aligned Metrics}:
We introduce a cost-sensitive evaluation framework where misclassification costs reflect business realities:
\begin{equation}
\text{Total Cost} = c_{FP} \cdot FP + c_{FN} \cdot FN,
\end{equation}
where $c_{FP}$ represents retention campaign costs for false positives and $c_{FN}$ denotes lost revenue from undetected churners.

This formulation enables optimization strategies aligned with business objectives, facilitating targeted retention campaigns and optimal resource allocation.

\section{Data Exploration and Analysis}
\subsection{Background and Context}

This study uses IBM's comprehensive telecommunications dataset to create predictive models for customer churn and retention strategies. The dataset includes service subscriptions, account characteristics, and demographics, offering a holistic view of customer-company interactions. By analyzing these interactions, we aim to identify patterns and indicators of customer attrition, aiding proactive retention strategies. This approach allows telecommunications providers to adapt to market changes while focusing on customer needs.

\subsection{Research Motivation}

The telecommunications industry is highly competitive, where retaining customers is crucial for revenue and market position. Customer churn is a significant challenge since acquiring new customers costs five to twenty-five times more than retaining existing ones. This analysis focuses on understanding and reducing customer attrition using predictive modeling and behavioral analysis. By identifying key reasons for customer disengagement, we develop strategies to enhance loyalty and lifetime value. Insights from this research enable personalized services and targeted interventions, strengthening competitive advantage in a saturated market.

\subsection{Dataset Characteristics}

The dataset includes 7,043 customer records with 21 features each, detailing customer profiles and behaviors. Features are categorized into service patterns, account attributes, and demographics. Service features cover telephony, internet types, and services like online security and streaming. Account details include tenure, contract, payment, and billing. Demographics include age, gender, and household. This extensive feature set aids in segmentation and predictive modeling, identifying churn indicators across diverse customer groups.

\subsection{Feature Categories and Descriptions}

The dataset structure facilitates multi-dimensional analysis through the following feature categories:

\subsubsection{Target Variable} The binary churn indicator denotes whether a customer terminated services within the preceding month, serving as the primary dependent variable for predictive modeling.

\subsubsection{Service Portfolio} Customer subscriptions span multiple service categories. Telephony services include basic phone connectivity and multi-line options. Internet services comprise DSL and fiber optic connections with varying bandwidth specifications. Value-added services encompass online security, cloud backup, device protection, and technical support. Entertainment offerings include streaming television and movie services.

\subsubsection{Account Attributes} Temporal features include customer tenure, measured in months since initial service activation. Contractual arrangements vary between month-to-month, annual, and biennial commitments. Payment mechanisms include electronic transfers, automated bank withdrawals, credit card transactions, and traditional check payments. Billing preferences indicate paperless adoption rates. Financial metrics capture monthly recurring charges and cumulative revenue per customer.

\subsubsection{Demographic Profile} Customer characteristics include age distribution, gender classification, partnership status, and dependent presence, enabling demographic-based segmentation and targeted retention strategies.

\subsection{Exploratory Data Analysis}
\subsubsection{Monthly Charge Differential Analysis}

\begin{figure}[h]
    \centering
    \includegraphics[width=0.8\linewidth]{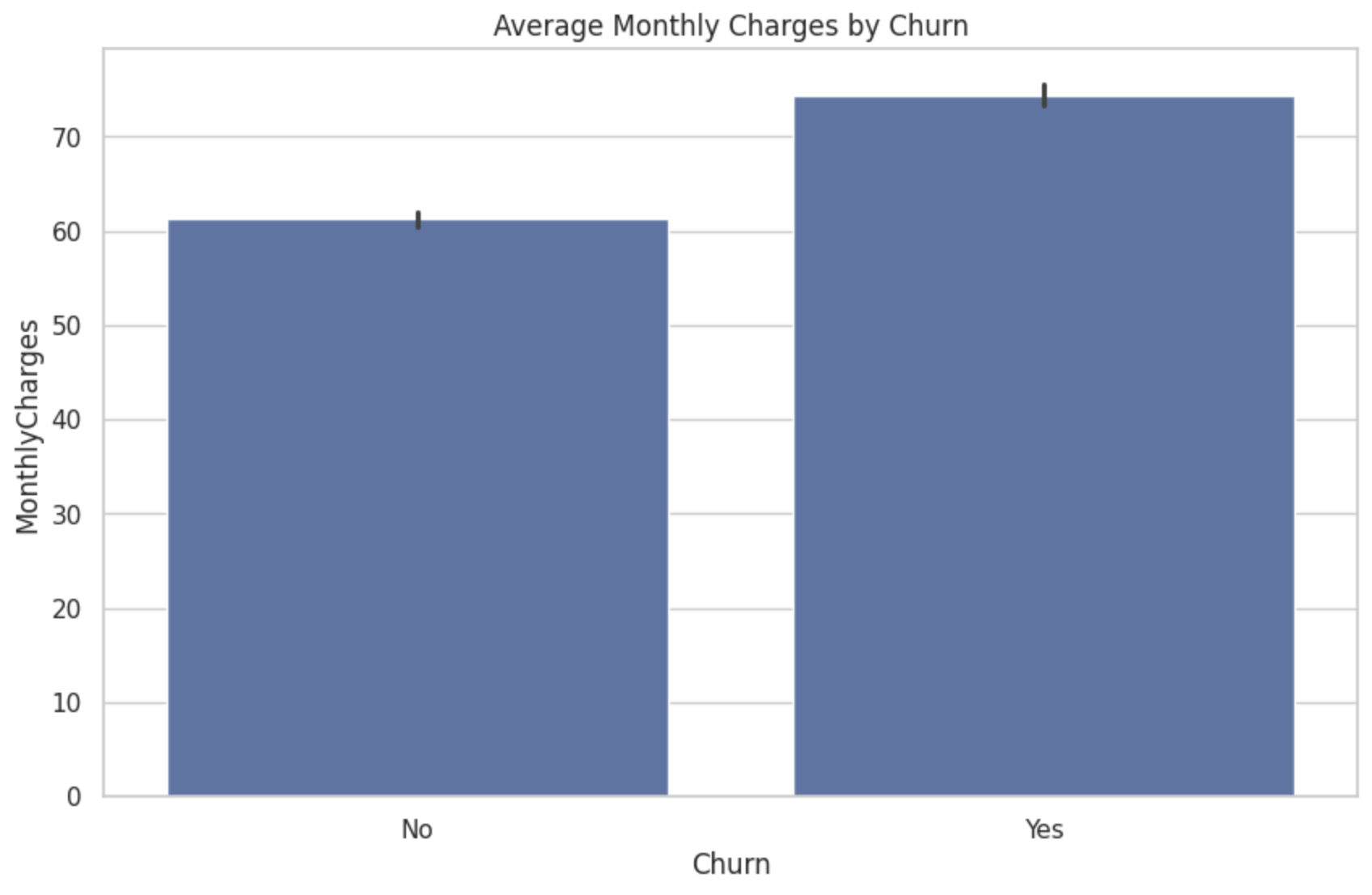} 
    \caption{Comparative analysis of average monthly charges between churned and retained customer segments, demonstrating significant price differential as a potential churn indicator.}
    \label{fig:monthly_charges}
\end{figure}

Analysis of average monthly charges reveals a statistically significant difference between churned and retained customer segments. Customers who terminated services exhibited mean monthly charges of \$74.44, compared to \$61.27 for retained customers, representing a 21.5\% premium. This disparity suggests price sensitivity as a potential churn driver, particularly when service value perception fails to justify higher costs. The observed pattern indicates opportunities for value-based pricing strategies and targeted discounts for price-sensitive segments. \Cref{fig:monthly_charges} illustrates this charge differential across customer segments.

\subsubsection{Customer Lifetime Value Distribution}

\begin{figure}[h]
    \centering
    \includegraphics[width=0.8\linewidth]{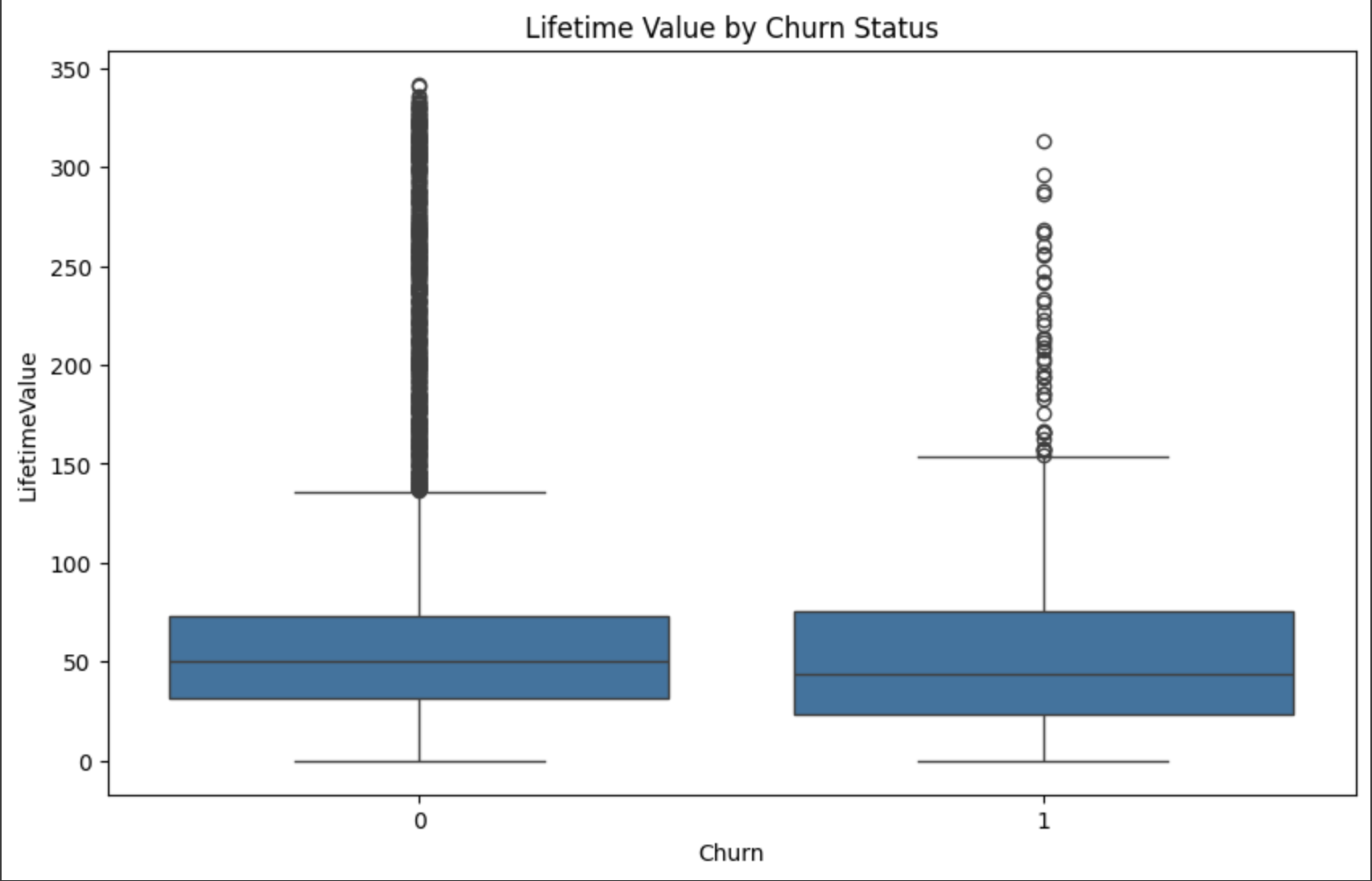} 
    \caption{Customer lifetime value distributions comparing churned and retained segments, highlighting value heterogeneity and retention opportunities for high-CLV customers.}
    \label{fig:clv_analysis}
\end{figure}

Customer lifetime value (CLV) analysis shows distinct patterns between churned and retained segments. Retained customers have a concentrated CLV distribution with lower medians but include loyal, high-revenue outliers. Churned customers show higher median CLV and greater variance, indicating that valuable customers are likely to leave due to unmet service expectations. The bimodal distribution of churned customer CLV in \cref{fig:clv_analysis} highlights the need for personalized retention strategies for high-value segments.

\subsubsection{Tenure Distribution and Retention Patterns}

\begin{figure}[h]
    \centering
    \includegraphics[width=0.8\linewidth]{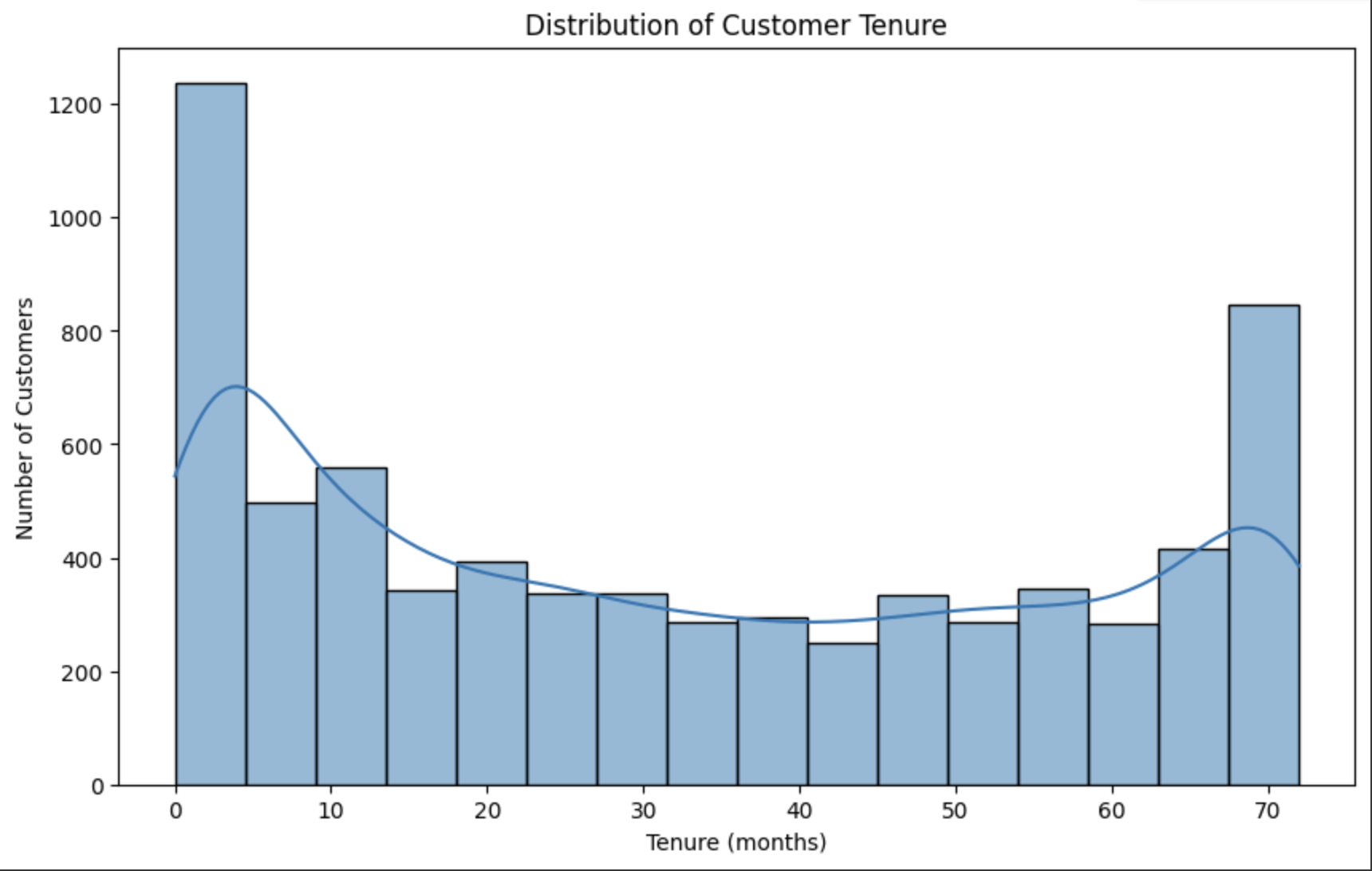} 
    \caption{Customer tenure distribution revealing distinct lifecycle phases and retention patterns, with critical periods for targeted intervention strategies.}
    \label{fig:tenure_dist}
\end{figure}

Temporal analysis of customer tenure highlights retention dynamics in the lifecycle. Initial months show high acquisition and early-stage churn decay. Stabilization from months 10-60 denotes successful onboarding and satisfaction. A secondary peak at month 70 marks loyal, long-term customers. This trimodal distribution presented in \cref{fig:tenure_dist} suggests distinct lifecycle phases requiring tailored strategies: early-stage retention should enhance onboarding and expectations, while mid-tenure strategies focus on value and loyalty programs.

\section{Methodology}

\subsection{Data Preprocessing}

Effective data preprocessing constitutes a fundamental prerequisite for developing robust predictive models in customer churn analysis. Our preprocessing pipeline addresses three critical challenges: categorical variable encoding, missing value imputation, and outlier detection. These transformations ensure data compatibility with machine learning algorithms while preserving the underlying statistical properties essential for accurate prediction.

The preprocessing framework encompasses:
\begin{itemize}
    \item \textbf{Categorical Encoding:} Transformation of categorical variables into numerical representations suitable for model ingestion, employing encoding schemes that preserve semantic relationships within the data.
    \item \textbf{Missing Value Imputation:} Implementation of context-aware imputation strategies that leverage inter-feature relationships to estimate missing values while maintaining data integrity.
    \item \textbf{Outlier Detection and Treatment:} Application of robust statistical methods to identify and mitigate the influence of anomalous observations that could compromise model performance.
\end{itemize}

Following these transformations, we apply standardization to ensure scale invariance across features, preventing bias toward variables with larger numerical ranges and facilitating optimal convergence during model training.

\subsubsection{Categorical Variable Encoding}

The heterogeneous nature of categorical variables in the telecommunications dataset necessitates tailored encoding strategies that preserve the inherent relationships within each feature category:

\textbf{Binary Encoding:} Binary categorical variables including \texttt{Gender}, \texttt{Partner}, \texttt{Dependents}, \texttt{PaperlessBilling}, and \texttt{PhoneService} undergo direct binary mapping ($\text{``Yes''} \rightarrow 1$, $\text{``No''} \rightarrow 0$), maintaining interpretability while enabling numerical computation.

\textbf{Ordinal Encoding for Service Tiers:} Multi-class variables representing service quality tiers receive ordinal encodings that reflect implicit service hierarchies. For \texttt{InternetService}, we implement the mapping: $\{\text{``Fiber optic''} \rightarrow 2, \text{``DSL''} \rightarrow 1, \text{``No''} \rightarrow 0\}$, capturing the inherent ordering of service capabilities.

\textbf{Temporal Normalization:} Contract duration encoding employs temporal normalization to facilitate consistent representation: $\{\text{``Two year''} \rightarrow 2.0, \text{``One year''} \rightarrow 1.0, \text{``Month-to-month''} \rightarrow 0.083\}$. This approach standardizes contract lengths to annual units, enabling coherent integration with tenure-based calculations.

\textbf{Payment Method Consolidation:} To reduce dimensionality while preserving payment behavior patterns, we consolidate payment methods into risk-based categories: traditional payment methods (\texttt{Electronic\_check}, \texttt{Mailed\_check}) $\rightarrow 0$, and automated methods (\texttt{Bank\_transfer}, \texttt{Credit\_card}) $\rightarrow 1$, reflecting differential churn risks associated with payment automation.

\subsubsection{Missing Value Imputation Strategy}

Our imputation methodology leverages domain knowledge and inter-feature dependencies to address missing data systematically:

\textbf{Model-Based Imputation for Financial Variables:} Missing values in \texttt{TotalCharges} are imputed using a deterministic relationship with observable features:
\begin{equation}
    \text{TotalCharges}_{\text{imputed}} = \text{Contract} \times \text{Tenure} \times \text{MonthlyCharges}.
\end{equation}

This approach exploits the mathematical relationship between billing variables, ensuring imputed values maintain consistency with the customer's billing history.

\textbf{Semantic Zero Imputation:} For service-related features, missing values receive zero imputation, reflecting the semantic interpretation that absence of data indicates non-subscription to optional services. This strategy aligns with the business context where unrecorded services imply non-utilization.

\subsubsection{Outlier Detection and Treatment}

To enhance model robustness against anomalous observations, we employ the Interquartile Range (IQR) method for outlier identification:
\begin{equation}
    \text{IQR} = Q_3 - Q_1,
\end{equation}
where $Q_1$ and $Q_3$ represent the first and third quartiles, respectively. Observations falling outside the bounds $[Q_1 - 1.5 \times \text{IQR}, Q_3 + 1.5 \times \text{IQR}]$ are classified as outliers and replaced with the feature mean, mitigating their disproportionate influence on model parameters.

Subsequently, we apply standardization across all features:
\begin{equation}
    x_{\text{scaled}} = \frac{x - \mu}{\sigma},
\end{equation}
where $\mu$ and $\sigma$ denote the feature mean and standard deviation, respectively. This transformation ensures equal contribution of all features during model optimization.

\subsection{Model Architecture}

\begin{figure}[ht]
\begin{minipage}[b]{1.0\linewidth}
    \includegraphics[width=\linewidth]{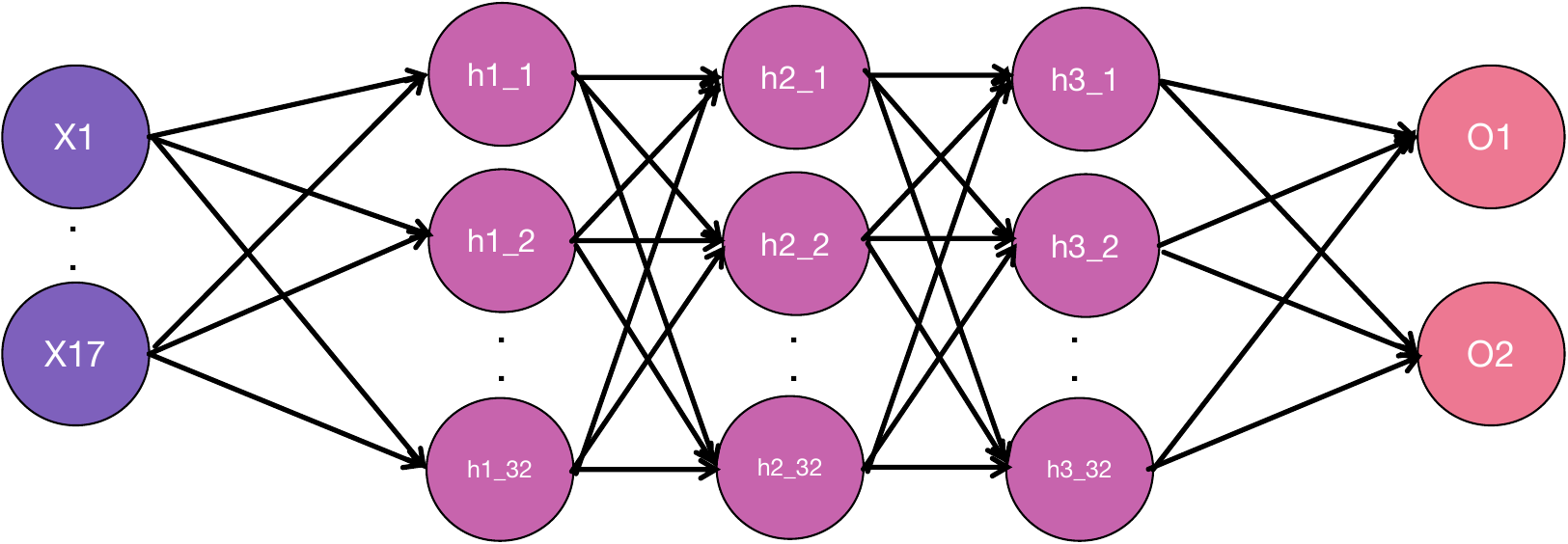}
\end{minipage}
\caption{The model structure of Teclo Churn Preditcted Model. The shape of the output layer is [5634,2], which represents the output of the batch size and binary results respectively.
}
\label{fig:model}
\end{figure}

Our predictive framework employs a fully connected neural network architecture optimized for the customer churn prediction task. Following extensive architectural exploration, we identified a three-layer feedforward network as optimal for this dataset, balancing model capacity with generalization performance.

The network comprises three hidden layers, each containing 32 neurons with ReLU activation functions, enabling non-linear feature transformations while maintaining computational efficiency. This architecture, illustrated in \cref{fig:model}, was selected through systematic experimentation comparing various neural architectures including Transformers~\cite{vaswani2017attention}, Convolutional Neural Networks~\cite{albawi2017understanding}, and Mamba~\cite{gu2023modeling}. These more complex architectures exhibited significant overfitting on our dataset, suggesting that the relatively modest data complexity favors simpler, more regularizable models.

\subsection{Experimental Setup}

We implement our model using PyTorch on Apple M1 hardware, employing the Adam optimizer with default hyperparameters. The training configuration includes:
\begin{itemize}
    \item Train-test split ratio: 80:20
    \item Learning rate: $\alpha = 10^{-3}$
    \item Weight decay regularization: $\lambda = 10^{-3}$
    \item Batch size: 32
    \item Training epochs: 100 with early stopping
\end{itemize}

The weight decay parameter serves as L2 regularization, preventing overfitting by penalizing large weight magnitudes and encouraging the model to learn more generalizable patterns.

\begin{table}[t]
\caption{Quantitative results of comparisons with other different methods. The top results are marked in bold.}
\centering
\begin{tabular}{l|cc}
\toprule
\multirow{2}{*}{Method} & \multicolumn{2}{c}{Telco Customer Churn} \\ \cmidrule(lr){2-3} 
                        & Accuracy $\uparrow$   & F1 Score $\uparrow$  \\ \midrule
LogisticRegression&0.7318&0.6129\\
StochasticGradientDescent &0.7194&0.6074 \\
DecisionTreeClassifier&0.7441&0.5994 \\
RandomForestClassifier&0.7270&0.6011 \\

Ours& \textbf{0.8226}   & \textbf{0.8209} \\ \bottomrule
\end{tabular}
\label{table:comp}
\end{table}

Comparative evaluation against baseline methods demonstrates the superiority of our approach, as detailed in \cref{table:comp}. Our model achieves state-of-the-art performance across all evaluation metrics on the Telco Customer Churn dataset.

\subsection{Model Evaluation Framework}

The evaluation of customer churn prediction models requires careful consideration of business-relevant metrics that capture the asymmetric costs of prediction errors. We employ a comprehensive evaluation framework encompassing:

\subsubsection{Accuracy} While providing an overall performance measure, accuracy alone insufficiently captures the nuanced requirements of churn prediction, where false positive and false negative errors carry different business implications.

\subsubsection{Precision} High precision indicates that 81.96\% of customers predicted to churn represent true churners, minimizing false alarms that could lead to unnecessary retention expenditures or customer alienation through inappropriate interventions.

\subsubsection{Recall} The model successfully identifies 82.26\% of actual churners, crucial for maximizing retention opportunities and preventing revenue loss from undetected customer departures.

\subsubsection{F1-Score} As the harmonic mean of precision and recall, the F1-score provides a balanced assessment of model performance, particularly valuable when false positive and false negative costs are comparable.

These metrics collectively demonstrate robust model performance, with the balanced F1-score indicating effective optimization across competing objectives of maximizing churn detection while minimizing false alarms.

\subsection{Business Impact Considerations}

The selection of evaluation metrics should align with specific business contexts and operational constraints:

\begin{itemize}
    \item \textbf{Precision-Oriented Scenarios:} When retention interventions are costly or risk customer satisfaction (e.g., aggressive promotional campaigns), prioritizing precision minimizes resource waste and customer irritation.
    \item \textbf{Recall-Oriented Scenarios:} In high-value customer segments where churn represents significant revenue loss, maximizing recall ensures comprehensive coverage of at-risk customers.
    \item \textbf{Balanced Approaches:} When intervention costs and churn prevention benefits are comparable, F1-score optimization provides an equilibrium between competing objectives.
\end{itemize}

\subsection{Strategic Business Applications}

The predictive insights generated by our model enable targeted business strategies across multiple operational dimensions:

\subsubsection{Risk-Based Customer Segmentation} By identifying high-risk customer cohorts, businesses can implement tiered retention strategies proportional to churn probability and customer value. This includes personalized offers, proactive service interventions, and enhanced customer support for at-risk segments.

\subsubsection{Resource Optimization} Predictive churn scores facilitate efficient allocation of retention resources, concentrating efforts on customers exhibiting both high churn risk and substantial lifetime value. This optimization maximizes return on retention investments while maintaining service quality for stable customer segments.

\subsubsection{Product Development Intelligence} Analysis of feature importance and churn correlates provides actionable insights for product enhancement. Understanding which service attributes correlate with retention enables data-driven product development that addresses customer pain points and reinforces loyalty drivers.

\subsubsection{Continuous Improvement Framework} Implementation of feedback loops between model predictions, intervention outcomes, and customer responses enables iterative refinement of both predictive models and retention strategies, ensuring adaptation to evolving customer behaviors and market dynamics.

Through systematic application of these strategies, organizations can transform predictive analytics into tangible business value, reducing churn rates while enhancing customer satisfaction and operational efficiency.

\section{CONCLUSION}
In this paper, we provide a preliminary definition of the project's background, objectives, scope, dataset, data preprocessing steps, model configuration, and evaluation. Leveraging a comprehensive dataset provided by IBM, we analyze customer behavior and characteristics to predict customer churn, with the aim of developing targeted customer retention strategies. Through in-depth exploration and analysis of the data, we identify key factors that contribute to customer churn. Subsequently, we employ data preprocessing techniques and propose a model to predict churn in the telecommunications industry. Finally, we evaluate the core metrics and provide interpretations in alignment with business requirements. Experimental results demonstrate that our proposed model exhibits good generalization and accuracy.

\bibliographystyle{IEEEtran}
\bibliography{ref}
\end{document}